\documentclass[prb,twocolumn,preprintnumbers,amsmath,amssymb]{revtex4}
\usepackage{graphicx}
\usepackage{dcolumn}
\usepackage{bm}
\usepackage{times}
\usepackage{pifont}
\usepackage{amsmath}
\usepackage{mathptmx}
\usepackage{color}

\begin{document}
\title{Symmetry-Tunable Environment in Functional Ferroic Heterostructures}

\author{C. Becher$^{1}$}
\author{M. Trassin$^{1,2}$}
\author{M. Lilienblum$^{1}$}
\author{C. T. Nelson$^{2}$}
\author{S. J. Suresha$^{2}$}
\author{D. Yi$^{2}$}
\author{P. Yu$^{2}$}
\author{R. Ramesh$^{2}$}
\author{M. Fiebig$^{1}$}
\author{D. Meier$^{1,2}$}
\email[]{dennis.meier@mat.ethz.ch}

\affiliation{$^{1}$Department of Materials, ETH Zurich, Wolfgang-Pauli-Strasse 10, 8093 Zurich,
Switzerland}

\affiliation{$^{2}$Department of Materials Science and Engineering, University of California,
Berkeley, CA-94720, USA}



\date{\today}

\begin{abstract}
We demonstrate a concept for post-growth symmetry control in oxide heterostructures. A functional
oxide is sandwiched between two ferroelectric layers and inversion symmetry at the position of the
oxide constituent is reversibly switched on or off by layer-selective electric-field poling. The
functionality of this process is monitored by optical second harmonic generation. The
generalization of our approach to other materials and symmetries is considered. We thus establish ferroic trilayer structures as device components in which
symmetry-driven charge-, spin-, and strain-related functionalities can be activated and
deactivated at will.
\end{abstract}

\maketitle

\begin{figure}[t]
    \centering
        \includegraphics[width=8.5cm,keepaspectratio,clip]{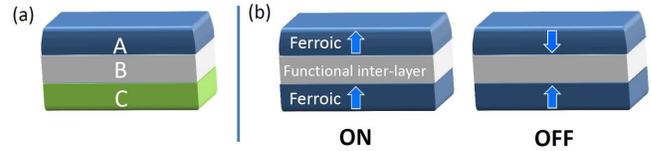}
\caption{(a) Classical ABC-trilayer heterostructure. (b) ABA-trilayer in which inversion symmetry
violation at the position of the functional B layer is actively set to ON or OFF by, respectively,
parallel or antiparallel orientation of the polarization as ferroic order parameter.}
    \label{fig:fig1}
\end{figure}

The correspondance of symmetry and functionality pervades all fields in physics and is a
cornerstone for optimizing the performance of functional materials for technological applications~\cite{Curie94,Livio12,Hwang12a}.
Symmetry violations in time and space, for instance, are a hallmark for the emergence of magnetic
and electric long-range order, respectively. If both of these symmetries are broken novel
interactions like magnetoelectric coupling effects are activated. A promising pathway towards
enhanced or novel functionalities is to incorporate such symmetry violations directly into the
material design. This becomes feasible when using high-end deposition techniques that allow for
symmetry engineering at the atomic level by growing atomically sharp interfaces~\cite{Hwang12a,Mannhart10a}.
Growth-induced inversion symmetry violation, for example, can activate spin-orbit mediated
phenomena such as spin-band splitting (Rashba effect~\cite{Rashba60a}) or spin canting
(Dzyaloshinskii-Moriya interaction~\cite{Moriya60a}). This is a source for spin-polarized currents
in thin-film semiconductors and ferromagnetic contributions in otherwise antiferromagnetic
systems~\cite{Koga02,Nitta97,Ederer05a,Cheong07}. A particularly elegant way for generating the desired symmetry reduction is acquired in
so-called tricolor superlattices where a geometric inversion symmetry breaking is achieved by the
stacking three of constituents, A - B - C, as illustrated in Fig.~\ref{fig:fig1}(a)~\cite{Sai00a,Warusawithana03a}. Here, as a consequence of the artificially
lowered symmetry, an enhanced electric polarization~\cite{Lee05a} and interface
magnetization~\cite{Yamada04a}, as well as unusual optical
properties~\cite{Zubko11a,Figotin03a,Lepri11a} may arise. Unfortunately, a growth-based
compositional symmetry breaking of the  ABC-type is inevitably immutable: It cannot be altered
after the completion of growth. In this sense, ABC-type heterostructures following the classical
device paradigm may be regarded as \textit{passive} building blocks. Here the step toward
\textit{active} heterostructures in which symmetry-driven functionalities can be repeatedly
switched on and off via reversible post-growth symmetry control would be a major achievement.

In this work we report such reversible post-growth symmetry control using epitaxial trilayer
heterostructures with controllable inversion-symmetry properties as model system. We sandwich a
functional candidate compound between two ferroelectric layers and violate or restore the net
inversion symmetry by controlling the associated polarization configuration as sketched in
Fig.~\ref{fig:fig1}(b). The latter is readily accomplished with moderate voltages applied at
room-temperature and monitored by optical second harmonic generation (SHG) as symmetry-sensitive
detection tool.

\begin{figure}[t]
    \centering
        \includegraphics[width=8.5cm,keepaspectratio,clip]{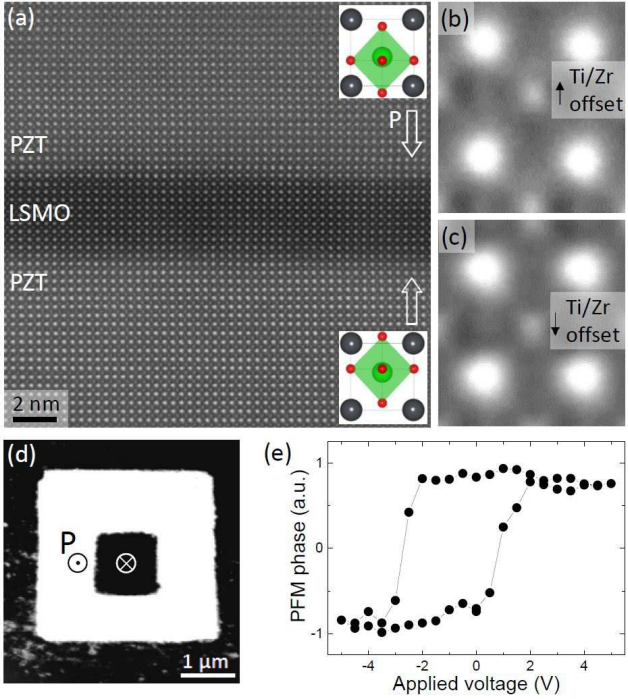}
\caption{(a) HAADF-STEM cross section image (Z-contrast) of a PZT/LSMO/PZT trilayer
(50~nm/5~nm/50~nm). Schematic insets in (a) illustrate the Zr/Ti cation displacements visible in
the blow-ups shown in (b) and (c). White arrows indicate the direction of the associated
spontaneous polarization. (d) Selective ferroelectric polarization switching in the upper PZT
layer. Out-of-plane PFM image of an electrically switched box-in-box area. (e) Corresponding
quasistatic piezoresponse hysteresis loop.}
    \label{fig:fig2}
\end{figure}

For realizing symmetry-tunable heterostructures we selected the textbook ferroelectric
PbZr$_{0.2}$Ti$_{0.8}$O$_3$ (PZT) as constituent providing the symmetry environment and metallic
La$_{0.7}$Sr$_{0.3}$MnO$_3$ (LSMO) as functional inter-layer. Epitaxial PZT(001) (50~nm) /LSMO(001) (5 nm) /PZT(001) (50~nm) trilayers were grown on
(001)-oriented SrTiO$_3$ (STO) substrates by pulsed laser deposition (PLD). The PZT thickness was
set to 50~nm in order to retain the epitaxial strain imposed by the substrate and thus maintain a
ferroelectric single-domain state~\cite{Nagarajan99a}. Figure~\ref{fig:fig2}(a) shows a cross
section of the PZT/LSMO/PZT trilayer system imaged by high-angle annular dark field (HAADF)
scanning transmission electron microscopy (STEM). The image (Z-contrast) reveals high-quality
interfaces without any evidence of inter-diffusion. Moreover, the direct visualization of the
structure allows to determine the trilayer polarization state which manifests as a cation
(Ti/Zr) off-centering along the [001] growth direction in the PZT unit cells. This offset is
visible in Figs.~\ref{fig:fig2}(b) and (c), which show a blow-up of the structure in the upper and
lower PZT layer of Fig.~\ref{fig:fig2}(a), respectively. Based on the observed Ti/Zr offset we
conclude that the polarization alternates according to $+P \leftrightarrow 0 \leftrightarrow -P$
across the LSMO inter-layer with a sharp transition at the PZT/LSMO interfaces.

Piezoresponse force microscopy (PFM) was applied in order to unveil the two trilayer
polarization states in the PZT/LSMO/PZT system illustrated in Fig.~\ref{fig:fig1}(b). We
investigated the effect of applying an electric voltage of $\pm 4$~V to the top ferroelectric
layer. Figure~\ref{fig:fig2}(d) shows the out-of-plane PFM contrast obtained after poling a
box-in-box structure with the PFM tip. Bright and dark regions correspond to a
ferroelectric single domain state with $+P$ and $-P$, respectively, clearly indicating that the upper layer
can be selectively and reversibly switched between two stable ferroelectric states. This
constitutes the two symmetry environments illustrated in Fig.~\ref{fig:fig1}(b). The locally
acquired hysteresis loop in the piezoelectric (resp.\ ferroelectric) response shown in
Fig.~\ref{fig:fig2}(e) reflects the excellent ferroelectric performance of the switchable PZT
control-layer. However, note that the negative bias of the local PFM loop in
Fig.~\ref{fig:fig2}(e) indicates an influence on the PZT switching behavior exerted by the
electrostatic environment or by interfacial strain~\cite{Yu12a,Vrejoiu08a}. Thus, although the PFM
data in Figs.~\ref{fig:fig2}(d) and (e) reveals the presence of two stables states for the upper
ferroelectric PZT layer, we cannot yet exclude that the lower PZT layer is at least partially
switched by the poling field.

For further verification of this issue we therefore devised an experiment that allows us to probe
the trilayer net symmetry in a convenient non-invasive and non-destructive way. Based on the
von-Neumann principle the gain and loss of inversion symmetry is expected to directly manifest in
the physical properties of the trilayer system, including its optical response. In particular,
optical frequency doubling --- also known as second harmonic generation (SHG) --- can, in the
leading electric-dipole order, only occur when the trilayer inversion symmetry is
broken~\cite{Fiebig04a,Denev11a}. This is described by the relation
\begin{equation}\label{eq:SHG}
 \vec P(2\omega)=\epsilon_0\hat\chi^{tri}\vec E(\omega)\vec E(\omega) \; ,
\end{equation}
with the electric field $\vec E(\omega)$ of the incident light wave at frequency $\omega$, the
induced electric-dipole oscillation $\vec P(2\omega)$ at frequency $2\omega$, and the effective
nonlinear susceptibility tensor $\hat\chi^{tri}$ of the trilayer. For an antiparallel orientation
of the polarization in the two PZT layers (OFF state in Fig.~\ref{fig:fig1}(b)) the systems
inversion symmetry is not violated so that $\hat\chi^{tri}\equiv 0$ holds. In contrast, inversion
symmetry is broken for parallel orientation of the polarization in the two PZT layers so that SHG
is symmetry-allowed in this ON state ($\hat\chi^{tri}\neq 0$). In order to confirm that SHG indeed
probes the entire trilayer net symmetry expressed by $\hat\chi^{tri}$, we measured a
model system that allows us to approach the geometry and symmetry of the full
trilayer system step by step.

\begin{figure}[t]
    \centering
        \includegraphics[width=8.5cm,keepaspectratio,clip]{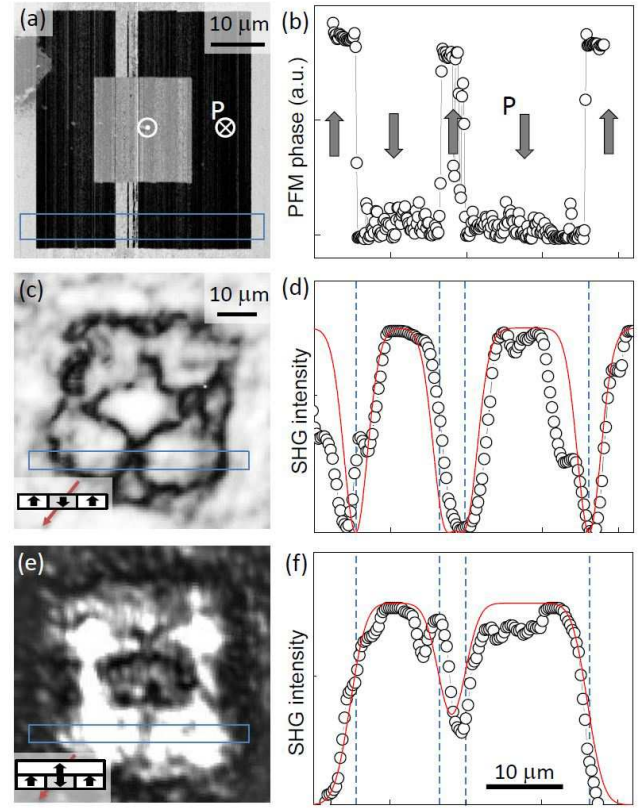}
\caption{(a) Out-of-plane PFM image of an electrically switched area in a PZT/SRO heterostructure.
(b) Line scan of the section marked by the blue box in (a). (c) SHG image of the of poled area shown in (a). A
schematic of the ferroelectric domain pattern is shown as inset with the red and black arrows indicating the
direction of the incident light and orientation of the ferroelectric polarization, respectively. (d) Line scan (black open dots) of the section marked in (c)
and calculated SHG intensity (red line). (e),(f) Corresponding SHG image, line scan, and
calculation for the ``construction kit'' trilayer system sketched in the inset (see text for
details).}
    \label{fig:fig3}
\end{figure}

At first, PZT single layers with a thickness of 50~nm were grown on SrRuO$_3$ (SRO)
buffered (001)-oriented STO. As indicated by the PFM image in Fig.~\ref{fig:fig3}(a) the PZT grows
in a ferroelectric single-domain state and can be reversibly switched by applying $\pm 4$~V to the
PFM tip. In Figs.~\ref{fig:fig3}(c) to (f) we show the corresponding SHG analysis performed with a
standard transmission setup described elsewhere~\cite{Fiebig04a}. In our experiments, light at a
photon energy around 1.0~eV was incident at an angle of about 20$^{\circ}$ with respect to the
surface normal of the PZT film. Because of the tilt, SHG components sensitive to the ferroelectric
polarization along the [001] direction can be excited. A spatially resolved SHG measurement taken
with SHG light from one of these components is presented in Fig.~\ref{fig:fig3}(c). The black
lines in the image retrace the ferroelectric 180$^{\circ}$ domain walls of the
electrically poled box-in-box structure. This is further highlighted in Figs.~\ref{fig:fig3}(b)
and (d) which show the variation of the PFM and SHG contrast along cross-sections of the
corresponding images. For modelling the lateral dependence $I(x)$ of the SHG intensity we
calculated the local interference of SHG contributions from neighboring domains with the
spontaneous polarization $+P$ and $-P$. Because of the linear coupling of the SHG susceptibility
to the spontaneous polarization the sign reversal of $P$ is converted into a $180^{\circ}$ phase
difference between SHG waves emitted from opposite domains~\cite{Fiebig02a}. Thus, the net SHG
yield is given by
\begin{equation}\label{eq:int}
 I(x) \propto \int sgn(P(x^{\prime}))exp\left(-\left(\frac{-(x-x^{\prime})}{l}\right)^2\right)dx^{\prime} \; .
\end{equation}
Here $sgn (P(x^{\prime}))$ accounts for the domain distribution in Fig.~\ref{fig:fig3}(a) and $l =
2.5$~$\mu$m is entered as optical resolution. The good agreement between the SHG data and the
calculated intensity profile in Fig.~\ref{fig:fig3}(d) corroborates that the SHG process is
sensitive to the polarization state of the single 50~nm PZT layer investigated here.

Taking advantage of this sensitivity we then investigated the SHG response of an assembly that
emulates the final symmetry-controlled ABA-type structures of Fig.~\ref{fig:fig1}(b) by a ``trilayer
construction kit''. This was built by mounting the switched PZT film of Fig.~\ref{fig:fig3}(a)
on top of a single-domain PZT/SRO/STO heterostructure with the two PZT layers facing each
other. Repeating the SHG experiments on this model trilayer we now obtained a pronounced contrast
in the SHG image shown in Fig.~\ref{fig:fig3}(e). Because of the well-defined polarization state
of the ``trilayer construction kit'' (see inset to Fig.~\ref{fig:fig3}(e)) the SHG brightness
levels can be unambiguously assigned to the local symmetry state. We find that dark regions correspond
to a centrosymmetric trilayer state, whereas bright regions are observed when inversion symmetry is broken~\cite{Kaneshiro08a}.
The good agreement between the SHG data and the calculated intensity profile in Fig.~\ref{fig:fig3}(f)
confirms that the SHG method is capable of probing the ferroelectric polarization state in buried
layers expressed by the coupling to $\hat\chi^{tri}$ and hence the net symmetry of the entire
arrangement.

\begin{figure}[t]
    \centering
        \includegraphics[width=8.5cm,keepaspectratio,clip]{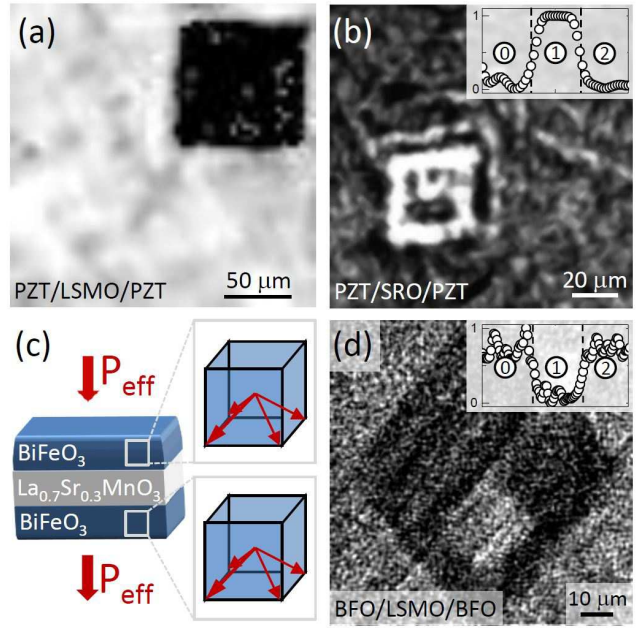}
\caption{(a) Spatially resolved SHG data taken on PZT/LSMO/PZT after electrically poling a box
with a PFM tip. (b) SHG image of an electrically poled box-in-box taken on PZT/SRO/PZT. As shown
in the inset, the SHG yield (vertical axis) can be reversibly switched between two discrete
levels. Numbers indicate the state at the start of the experiment (0), as well as the
states obtained after the first (1) and second (2) polarization reversal. (c)
Schematic illustration of the ferroelectric domain structure in a BFO/LSMO/BFO trilayer system in the ON state.
(d) Corresponding spatially resolved SHG data imaged after poling a box-in-box. The reversibility
of the switch is depicted in the inset showing the SHG yield obtained in states (0),(1), and (2).}
    \label{fig:fig4}
\end{figure}

The empirical demonstration leads us to the final step: the SHG analysis of the real trilayer
system, i.e.\ on the epitaxial PZT/LSMO/PZT trilayer grown on a SRO-buffered STO substrate.
Figure~\ref{fig:fig4}(a) displays the corresponding spatially resolved SHG data taken after
electrically poling an area of $80 \times 80$~$\mu$m$^2$ with a PFM tip. The poled area (dark)
shows a distinct contrast level and is clearly distinguished from the surrounding region (bright).
Referring to Fig.~\ref{fig:fig3}(e) we can conclude that the pronounced SHG contrast indicates a
polarization reversal in the top PZT layer, and along with it a symmetry change in the PZT/LSMO/PZT trilayer. In the present case the system has been
switched from a non-centrosymmetic (white $\leftrightarrow$ ON) to a centrosymmetric (black
$\leftrightarrow$ OFF) state. This switching behavior confirms that the top and bottom control
layers are decoupled which is essential for manipulating the symmetry state at the position of the
functional layer. In the present case the decoupling is supported by conduction in the LSMO
inter-layer. The inter-layer, however, does not have to be conducting: the decoupling of the
control layers may as well be acquired via bias effects as seen in Fig.~\ref{fig:fig2}(b) or by
pinning the polarization of the lower control layer, e.g. by doping or an additional bottom
electrode.

Note that the concept for active inversion symmetry control by ferroelectric poling is not
restricted to the case of PZT/LSMO/PZT trilayers. As suggested by Fig.~\ref{fig:fig1}(b) it is a
universal approach in the sense that no restrictions apply regarding the materials which are
chosen as the functional inter-layer and the ferroelectric control-layers. In order to verify the
general validity we realized different trilayer compositions by exchanging, in two separate
variations, either the material used for (i) the functional inter-layer or (ii) the ferroelectric
control-layers.

(i) We begin with a discussion of the effects observed when replacing LSMO for SRO. The PFM
analysis on PLD grown high-quality PZT/SRO/PZT trilayers on STO (001) revealed that the top PZT
layer grows in a ferroelectric mono-domain state and can be reversibly switched by applying a
voltage of $\pm 4$~V to the PFM tip (not shown). In Fig.~\ref{fig:fig4}(b) we present the
corresponding spatially resolved SHG data taken on a PZT/SRO/PZT trilayer (50~nm/25~nm/50~nm)
after electrically poling a box-in-box structure. In the SHG image poled and unpoled areas are
again clearly distinguishable due to pronounced SHG contrasts that, analogous to the case of LSMO,
relate to the local symmetry state of the system (dark $\leftrightarrow$ centrosymmetric, bright
$\leftrightarrow$ non-centrosymmetric). This evidences that the substitution of the inter-layer
material has no fundamental influence on the basic trilayer performance of Fig.~\ref{fig:fig1}(b) so
that we have the freedom to choose the inter-layer material according to the desired
functionality.

(ii) In the next step we replaced the ferroelectric PZT control-layers by multiferroic BiFeO$_3$
(BFO) films while keeping the initial LSMO inter-layer. In BFO the coexistence of electric and
magnetic order simultaneously violates space- and time-reversal symmetry. Thus, the use of BFO as
control-layer constituent allows us to exploit the electronic \textit{and} the spin degrees of
freedom in our symmetry-tunable trilayer environment. BFO/LSMO/BFO trilayers were grown on
(001)-oriented STO substrates leading to four ferroelectric (multiferroic) in-plane variants with
$P||\left\langle 111\right\rangle$ (see Fig.~\ref{fig:fig4}(c)) resulting from the cubic symmetry of the
substrate~\cite{Chu07a,Yu12a}. The associated domain sizes, however, are in the order of
10--100~nm. It is therefore sufficient to consider only the effective out-of-plane polarization
component, $P_{\text{eff}}$, for analyzing the symmetry state that is seen by SHG and other
mesoscopic probes with a limited resolution in the micrometer-regime. From previous work it is further
known that $P_{\text{eff}}$ is accompanied by a collinearly oriented small spontaneous
magnetization~\cite{Trassin13a}, $M_{\text{eff}}$, which results from a canting in the
antiferromagnetic spin arrangement of the BFO films~\cite{Ederer05a}. Thus, any reversal of
$P_{\text{eff}}$ coincides with a reorientation of $M_{\text{eff}}$ which extends the symmetry
control in our trilayers toward time reversal and the spin system.

PFM scans performed on a BFO/LSMO/BFO trilayer (50~nm/30~nm/50~nm) revealed a homogenous
out-of-plane contrast for the upper BFO control-layer (see Fig.~\ref{fig:fig4}(c) for a schematic
illustration). Figure~\ref{fig:fig4}(d) presents the corresponding SHG data taken after
electrically poling a box-in-box structure with a PFM tip ($\pm 10$~V). Similar to
Fig.~\ref{fig:fig4}(a) a striking decrease in the SHG yield is observed when switching
$P_{\text{eff}}$ ($\leftrightarrow M_{\text{eff}}$) in the upper BFO control-layer. It shows that
the poling process establishes space- and time-reversal symmetry which,
prior to the poling, have been broken in the BFO/LSMO/BFO system. A second electrical switch (inner box) then
leads to full recovery of the initial signal strength which demonstrates that the symmetry
violations in space and time are fully reversible.

In conclusion we demonstrated a concept for post-growth symmetry control in thin film
heterostructures consisting of a functional constituent sandwiched between two ferroic control
layers. As model case we discussed the activation and deactivation of a non-centrosymmetric
environment that was achieved by selective voltage-driven switching in one out of two
ferroelectric control layers. Subsequent experiments demonstrated the generality of the approach.
It is robust against an exchange of the control as well as the functional layers. By involving
multiferroic BFO we showed that even the type of environmental control is selectable. Aside from the
spatial inversion exemplarily discussed in this work it may involve space-time reversal via multiferroic
control layers (extending the classical spin-valve geometry~\cite{Chappert07a}), or strain via ferroelastic control layers. Our concept thus reveals a
universal route for gaining control of electric, magnetic, distortive, or even orbital degrees of
freedom post-growth which is a basis for applications employing heterostructure as \textit{active}
device components in which tailored asymmetry-driven functionalities can be switched on and off at
will.

M.T. acknowledges subsidy from the Center for Energy Efficient Electronics Science (NSF Grant No.
0939514). D.M. appreciates support by the Alexander von Humboldt Foundation. The authors thank J.
Seidel, J. Heron, and J. Clarkson for experimental assistance.

\end{document}